# MARKING NOON: THE TIME BALLS AND TIME FLAPS OF THE NETHERLANDS

**Richard de Grijs**
*School of Mathematical and Physical Sciences, Macquarie University,
Balaclava Road, Sydney, NSW 2109, Australia*
Email: richard.de-grijs@mq.edu.au

**Abstract:** In the nineteenth century, the Netherlands quickly adopted the time ball—a British innovation for maritime chronometer calibration—in its main naval ports (Nieuwediep/Den Helder, Vlissingen, Hellevoetsluis) and commercial centres (Amsterdam, Rotterdam). A large sphere dropped from a mast at a fixed time, the device enabled ships to verify their chronometers against a standard, essential for accurate longitude determination and safe navigation. Its ready acceptance was eased by indigenous Dutch traditions. Rural communities had long used visual time signals like the *sjouw* on Terschelling island, a wicker ball raised on a mast to mark the lunch hour and milking time for farmers, and the *lawei*, a basket or sack used in the peat bogs of Friesland to regulate labourers' hours. The Dutch time-signal system was distinguished by its strong institutional backing from the country's Royal Navy, its Hydrographic Service and by professional astronomers. Among the latter, Frederik Kaiser was a pivotal figure, vehemently defending the system's accuracy and pioneering technical improvements. He successfully advocated for replacing the traditional falling ball with a system of rotating flaps, which provided a more instantaneous and reliable visual signal. Beyond their practical role, time signals became civic spectacles and symbols of Dutch scientific modernity and imperial reach. Their decline began with the electric telegraph and was finalised by wireless radio, which allowed ships to calibrate chronometers at sea.

## 1 SETTING THE STAGE

Until well into the nineteenth century, seafaring nations faced one of the most pressing technical problems of modern navigation: the reliable determination of longitude at sea. Although longitude principles were understood by the seventeenth century (e.g., de Grijs, 2017), practical solutions depended on two technologies. The first was the development of precision chronometers that could maintain accurate time during long voyages under variable climatic and mechanical conditions. The second was the establishment of dependable reference signals on land, against which such chronometers could be calibrated.

In Great Britain, these requirements led to the invention of the time ball, a large painted wooden or metal sphere hoisted on a mast and dropped at a precisely defined moment each day to provide ships in port with a public time signal. Time balls were mounted high for visibility, usually atop towers or lighthouses. The poles that carried the balls were often topped with a weather vane and a directional cross indicating the four cardinal points.

### 1.1 An emerging concept

The concept of time balls "… for communicating time by means of telegraphs …" was first proposed in 1818 by Robert Wauchope (1788–1862), a Royal Navy officer (Bartky and Dick, 1981). Wauchope envisioned that "… there will … be no port of any consequence … where an accurate [going] rate for timepieces … may not be found" (*ibid.*: 158). He eventually published his practical manual, 'Time signals for Chronometers', in the *Nautical Magazine* in 1836 (Wauchope, 1836). In that same magazine issue, the editors pointed out what Wauchope's concept would evolve into:

> The plan there proposed of giving the instant of time, is by the observer with the sextant and artificial horizon on shore, causing at the instant of observation the dropping of a shutter on its hinges, the instant of its fall being noted on board as shewn by the time-keeper. (*Nautical Magazine*, v, 463; 1836).

Despite his advocacy, it took until 1829 before the Admiralty agreed to establish Britain's first operational time ball at Portsmouth (e.g., Kinns et al., 2021). This was followed by time balls in Mauritius and at the Royal Observatory, Greenwich, in 1833, at St. Helena in 1834 and at the Royal Observatory in Cape Town in 1836 (Kinns, 2022).

Time balls spread rapidly across Britain, Europe and beyond, soon becoming indispensable for navigators and port authorities. By the mid-nineteenth century, they had become a prominent feature in almost every significant maritime centre. Time balls had been established as far afield as, e.g., Calcutta (Kolkata, India; 1835), Batavia/Jakarta (Indonesia; 1839), Cape Coast Castle (Ghana; 1839), Bombay (Mumbai, India; 1840), Valparaiso (Chile; 1844), Liverpool (1845) and—following the founding of the U.S. Naval Observatory—Washington, D.C. (1845; Bartky and Dick, 1982). (For a comprehensive discussion of historical time signals, see Kinns, 2022.) Their purpose was simple yet transformative: by allowing ships' officers to check their chronometers against a known reference time—usually Greenwich Mean Time (GMT)—they enabled more accurate charting and safer global navigation.

The Kingdom of the Netherlands, long a maritime power with global commercial and colonial interests, was quick to recognise the benefits of this innovation. Dutch ports such as Amsterdam, Rotterdam, Vlissingen, Hellevoetsluis and Nieuwediep (Den Helder) installed their own time balls in the mid-nineteenth century, integrating them into a wider system of naval hydrography and astronomical observation networks.

### 1.2 Time balls as global technology

Although the historiography of navigation has long emphasised the British role in the global diffusion of precision timekeeping, the Dutch case deserves more than a cursory reference. For one, the Netherlands was at the heart of nineteenth-century European shipping networks, with the port of Rotterdam rapidly emerging as a central node in global trade. Moreover, Dutch adoption of time balls was not mere technological borrowing but part of a broader institutional culture in which the *Koninklijke Marine* (the Dutch Royal Navy), the *Hydrographisch Bureau* (Dutch Hydrographic Service) and the country's professional astronomers collaborated to integrate standards of measurement. Dutch time balls thus became more than navigational aids: they were instruments of scientific authority, civic spectacle and, in the Dutch colonial context, symbols of imperial power.

Their emergence coincided with nineteenth-century efforts to standardise time across increasingly interconnected societies (e.g., Barrows, 2011; Bartky, 2007). In Great Britain, France, Germany and the United States, observatories played a central role in producing reliable time signals, which were then disseminated to ports, railways and, eventually, the public at large. The visual drama of a large ball dropping from a mast, pole or tower offered both practical application and symbolic resonance: it embodied the precision of modern science in a form visible to all.

The Dutch adoption illustrates the transnational spread of such practices. Like their British counterparts, Dutch maritime authorities recognised that time signals could simultaneously serve multiple constituencies: seafarers in need of chronometer calibration, port officials regulating harbour activity and urban populations eager for a daily spectacle (and a time check). In Amsterdam, for example, the time ball was integrated into the city's astronomical infrastructure (see Section 3.1), while in Den Helder and Vlissingen (Sections 3.2 and 3.3, respectively) the navy base took responsibility for the maintenance and operation of the time signals. Rotterdam's time ball reflected the needs of a bustling commercial port, where merchant shipping predominated.

Here, I focus on Dutch time balls in Europe. I trace the rise, operation and eventual decline of time balls and time flaps in Dutch ports from their introduction in the 1830s and 1840s to their replacement by wireless time signals in the early twentieth century. In doing so, I attempt to situate Dutch practice within the wider European context, while also highlighting distinctively local features. I argue that the Dutch experience with time balls illustrates the intersection of global scientific practice, national maritime institutions and local cultural traditions.

## 2 EARLY DUTCH TIME SIGNALS

The time ball, as it emerged in Britain in the 1820s and 1830s, may seem at first to have been an entirely novel invention, the product of naval ingenuity and astronomical precision. Yet in the Netherlands, as elsewhere in Europe, the idea of an elevated public time signal was not unfamiliar. Long before Dutch harbours saw the daily rise and fall of time balls and other time signals, rural communities had relied on analogous devices to regulate the rhythms of work and daily life. These earlier practices form a cultural backdrop to the adoption of maritime time balls.

### 2.1 Early precursors

One of the most striking examples of such indigenous time signals is the *sjouw* of the northern Dutch island of Terschelling.[1] The *sjouw* was essentially a large wicker ball, fashioned of reeds or osiers (willow branches), which was hoisted each day on a tall mast, often a salvaged ship's spar repurposed for village use. Unlike the maritime time ball, whose exact timing was dictated by astronomical observation, the *sjouw* functioned as a pragmatic, communal signal for the structuring of agricultural labour.

Each morning at 11:30, the local schoolmaster—or, more likely, one of his pupils—raised the ball to its full height, visible across the fields and from neighbouring villages. The schoolmaster's contract included the explicit clause that he had to be able to tell the time (van der Giesen, 2017). This implies that the local church must have had an operational public clock as well.

The raised ball signalled to farmers and labourers—few of whom owned watches or could even tell the time—that the midday meal was near. Lunch would normally have been served at noon. Later in the day, at around 15:30, the ball was lowered, signalling to dairy farmers that it was time to bring in their cows or return to their farms themselves for milking. The *sjouw* was thus a shared temporal marker.

Although unconnected to astronomy or navigation, the *sjouw* embodied communal time discipline. Long before the mechanical clock or the railway timetable had imposed standardised hours across Dutch society, the *sjouw* provided a simple yet effective means of synchronising daily activity. It illustrates that Dutch communities were accustomed to the idea that time could be collectively marked by a publicly visible artefact.

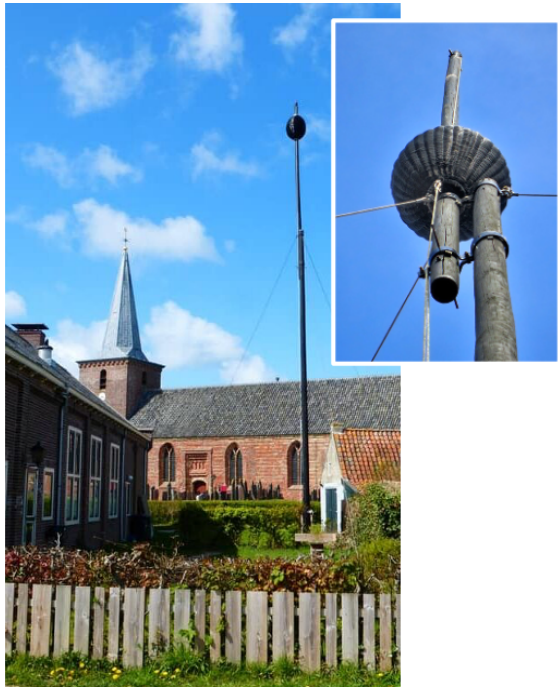

**Figure 1.** Sjouw on display in the Terschelling village of Hoorn, located next to the village church. (Aryan Westera, Mandvlechterij; reproduced with permission.)

Even in the twentieth century, traces of this tradition persisted. A *sjouw* has been preserved and can still be seen today in the village of Hoorn (Terschelling): see Figure 1. Located next to the thirteenth-century Sint Janskerk (St John's Church), it is displayed and operated as proud part of local heritage; it is still operated today between mid-June and mid-September (Naar Terschelling, 2025). Its survival underscores the cultural resonance of the practice, which, although humble in origin, offered a visual continuity between rural life and later technological innovations. Hoorn was originally an agricultural village, with livestock as its main source of income.

The *sjouw* in Hoorn is a ball-shaped, black-painted wicker basket that can be hoisted up and lowered around a pole. The pole in Hoorn is 18.4 metres high and consists of a lower mast with an extension on top, the topmast. This construction was commonly used in sea-going sailing vessels. The pole is kept upright by stays attached to the lower mast at the point where the topmast is fixed to it. This allows the wicker basket to be raised and lowered freely (van der Giesen, 2017).

Originally, a second *sjouw* was operated in the village of Midsland, further west on the island, so that both *sjouws* would have been easily visible from the entire agricultural area between the Wadden Sea (the 'Dutch Shallows') and the dunes protecting the island from the North Sea (van der Giesen, 2017). This became a serious challenge, however, when the local government undertook a sustained effort at tree planting across the island, starting around 1910.

Comparable practices existed elsewhere in the Netherlands as early as the seventeenth century.[2] In the peat-cutting districts of the northern province of Friesland, labourers employed a device known as a *lawei* (see Figure 2), essentially an empty (potato) basket, a sack or a bucket hoisted onto and lowered from a branch of a prominent tree to signal the working hours for the peat cutters (e.g., Goslinga, 2003; van der Molen, 1978). These signals, operated and supervised by the 'levay master' regulated communal labour in the peat bogs, ensuring that groups of workers, often spread over large expanses, could coordinate their activities. *Laweis* were intimately associated with peat extraction, to the extent that contracts issued to work the peatlands eventually became known as 'Laweys contracts'. Not only in peatlands, but also in land reclamation or dike construction, signal poles resembling *laweis* were used.

Like the *sjouw*, the *lawei* shows that Dutch rural communities had their own visual time-signalling methods. Both devices underscore a key point: time signalling by hoisted objects was a familiar cultural practice in the Netherlands long before the introduction of scientifically calibrated time balls in maritime ports.

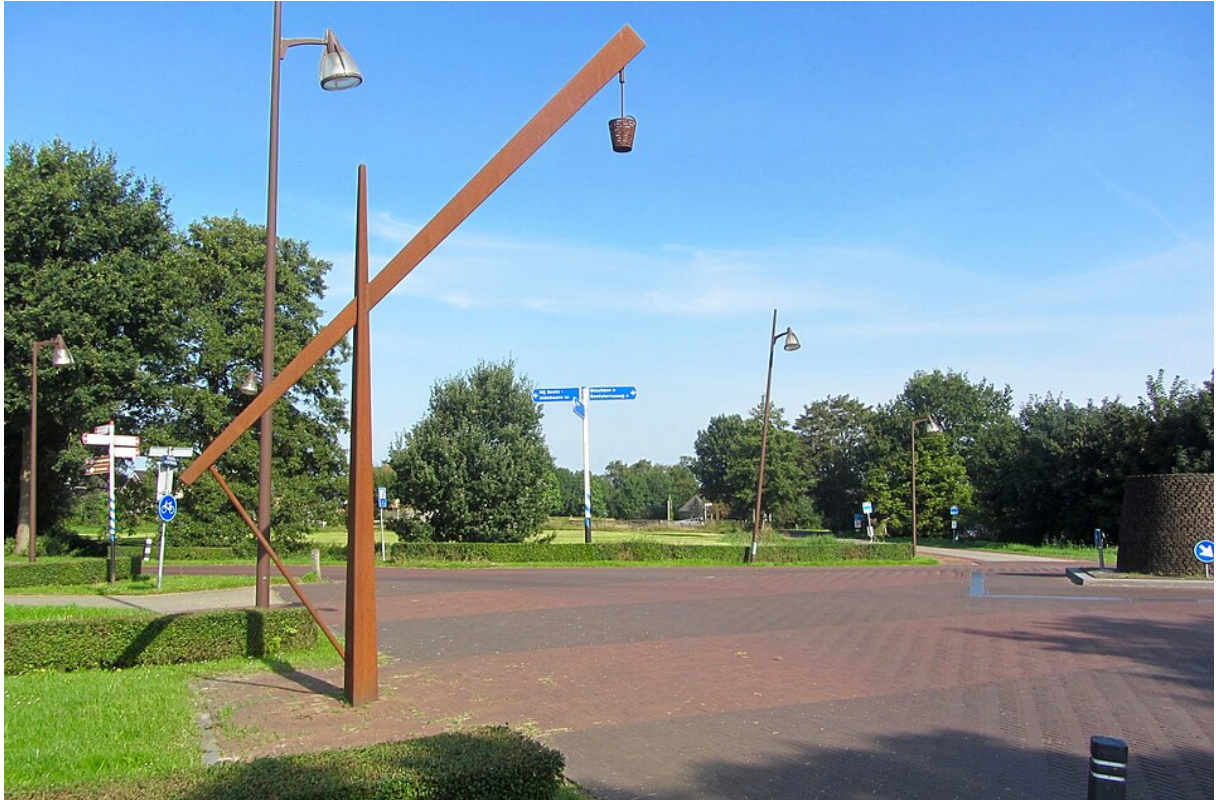
**Figure 2.** Modern *lawei* in the Frisian village of Nij Beets. (Anne Annema, 2008; CC BY-SA 4.0.)

## 2.2 Maritime visual signalling

Beyond rural labour, Dutch mariners and harbour communities were already accustomed to visual signalling for purposes other than time. Semaphore masts, flag systems and gunfire were routinely employed to mark arrivals, departures and warnings along the coast. The idea that ships' crews might look to a raised or lowered object onshore for critical information was therefore not alien.

The time ball was therefore less a disruption than a reconfiguration of existing practices within a scientific framework. What was novel was not the act of hoisting and dropping a visible object, but the precision and authority with which that act was timed. The crucial shift lay in linking the visible signal directly to astronomical observation and, by extension, to the determination of longitude.

When Wauchope first proposed his time-ball system to the British Admiralty in the late 1810s, he explicitly envisioned it as an extension of existing maritime signal traditions. The first successful installation at Portsmouth in 1829 (although deviating from Wauchope's design), and especially the Greenwich time ball in 1833, quickly demonstrated the benefits of the device for chronometer calibration. News travelled rapidly through naval and hydrographic circles, reaching Dutch authorities within a few years.

By the mid-1830s, Dutch hydrographic journals such as the *Tijdschrift toegewijd aan het Zeewezen* ('Journal devoted to the Maritime Service') and the *Verhandelingen en Berigten betrekkelijk het Zeewezen* ('Transactions and Reports on the Maritime Service') reported British practices and urged their adoption. The first proposal to establish a time ball in a Dutch port was submitted in 1838 to the *Maatschappij tot Nut der Zeevaart* ('Society for the Benefit of Navigation') in Rotterdam. Unfortunately, the costs associated with the rather ambitious plan were so high that it was not supported (Dekker, 1990).

A more modest scheme was approved by the Society in 1845, but before it could be implemented, the initiative to set up time balls had been taken over by naval officers, and the Society's plan threatened to be shelved indefinitely (Algemeen Rijksarchief, 1845; van Galen, 1846: 12–13). The first confirmation that time balls had been established in the Dutch sea ports of Willemsoord (Den Helder) and Hellevoetsluis was published in the *Staatscourant* (the Dutch Government Gazette) of 30 January 1846 (own translation):

<div style="text-align: center">MINISTRY OF THE NAVY.
NOTICE TO MARINERS.</div>

*Establishment of Time Balls at the Nieuwe Diep and at Hellevoetsluis.*

The Minister of the Navy
Hereby informs those concerned that, in order to provide the masters of ships bound for sea from *Hellevoetsluis* and from the *Texel Roads* with the opportunity, before sailing, to check and determine the time and rate of their timepieces with accuracy and convenience, the following provisional arrangement has been put in place—pending further measures—and is set to enter into operation on the *1st of February* next.

From one of His Majesty's ships, either at the *Nieuwe Diep*, at the *Texel Roads*, or in the harbour or at the *Hellevoetsluis Roads*, which will for this purpose be expressly provided with accurate timepieces, a time ball of about one Dutch ell in diameter will be dropped each day at precisely *noon, mean time of the place*, from a height between the topgallant yard and the topsail yard.

This time ball will be hoisted to *half-mast* precisely *five minutes* before local mean noon, as a warning for observers to prepare for the observation, and *two minutes* before that time it will be raised to its highest position.

The moment the ball *begins* to fall shall be taken as the instant of local mean noon.

The time difference in longitude between the positions of the ships entrusted with this service in the above-mentioned harbours and the meridian of the *Greenwich* Observatory is:

for the *Nieuwe Diep*: . . . . . . . . .  19′  7″
for the harbour of *Hellevoetsluis*: 16′ 32″

Furthermore, the masters of vessels arriving only in the afternoon at the aforementioned sea stations, and wishing to sail either the same day or the following morning, may apply to the commander of the ship entrusted with the time ball for a special opportunity to check their timepieces against the time ball.

The Hague, 28 January 1846
*The Minister aforesaid,*
J[ulius]. C[onstantijn]. Rijk.

While the Dutch installations of the 1840s were undoubtedly inspired by British models, they were received within a cultural landscape already accustomed to the visible marking of time. This helps explain the rapid acceptance of time balls in multiple Dutch ports: what might have seemed an alien practice in other contexts resonated with older communal traditions. By contrast, the first German time ball, erected in Cuxhaven, was installed only in 1875 (Kinns, 2022).

Moreover, the Dutch examples highlight the ways in which time balls operated at the intersection of science and culture. Historians of technology often emphasise the radical novelty of nineteenth-century innovations, yet the Dutch case reminds us that such innovations often built upon long-standing practices. The maritime time ball was indeed new in its precision and global implications, but its basic mode of operation—hoisting and dropping a ball—was already embedded in Dutch cultural experience. This insight matters for two reasons. First, it helps explain why Dutch ports embraced time balls without significant controversy: they were recognisable extensions of existing practices rather than entirely unfamiliar impositions. Second, it situates the Dutch contribution within a broader European history of time signalling, showing how global technologies were always locally inflected. The time ball, far from being a purely foreign innovation, thus represented a synthesis of international science and local tradition.

## 3 THE MAIN DUTCH MARITIME TIMEBALLS

The introduction of time balls into the Netherlands was both rapid and geographically extensive. Within two decades of Wauchope's first experiments at Portsmouth and Greenwich, Dutch ports from Den Helder in the north to Vlissingen (Flushing) in the south had installed their own devices. Each served as a local node in a broader network of maritime time signalling, collectively integrating the Netherlands into the global regime of precision timekeeping. In this section I reconstruct the geography, operation and personnel of the Dutch time-ball system, highlighting both its technical details and its social significance.

## 3.1 Amsterdam

Amsterdam, long the commercial heart of the Netherlands, was among the earliest Dutch cities to adopt a public time signal. As a major departure point, Amsterdam required reliable chronometer calibration: an error of four seconds meant one nautical mile at the equator.

By the 1840s, a daily time ball served shipping on the city's waterfront, the IJ (pronounced [ɛi]), closely linked to the city's astronomical and meteorological infrastructure. Hydrographic listings record that the Amsterdam signal was dropped at 11h 40m 21s 'mean time Greenwich' (Groustra, 1888: 161), a precision that reflects the exact longitude correction of the port and allowed mariners to reconcile their chronometers directly with Greenwich-based calculations. Note that in this article I have adopted references to 'common' time throughout. In contemporary manuals, this was often offset by twelve hours, as exemplified by the following instruction from 1853:

> The time ball … is hoisted halfway up the mast 5 minutes before local mean noon; 2 minutes before noon it is raised to the top, and it drops at 0h 0m 0s local mean noon, which corresponds to 23h 40m 53s mean time of the preceding day at Greenwich. (Hondeijker, 1853: 171)

The signal's location overlooking the IJ made it highly visible. According to contemporary standard practice, the ball was hoisted to half-mast several minutes before the appointed time, raised fully shortly before noon and then dropped exactly on the signal. On board, the cry *"bal neer!"* ("ball down!") alerted the officer of the watch to mark the difference between shipboard chronometers and the official signal. Repeated daily, this ritual gave captains and officers a record of their chronometer's *going rate*, that is, the rate of a chronometer's drift, essential for long-distance navigation (e.g., Dekker, 1990; Tooth, 1992). The routine became so ingrained that the phrase "*Het is vijf voor twaalf*" ('The time is five to twelve') became idiomatic (e.g., Vonkenberg, 1922–1923), similar in its meaning as the English expression 'the eleventh hour'.

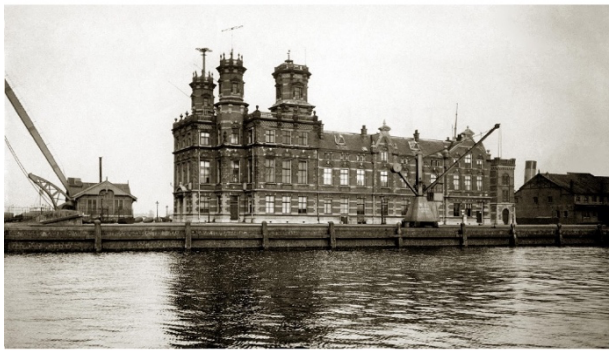

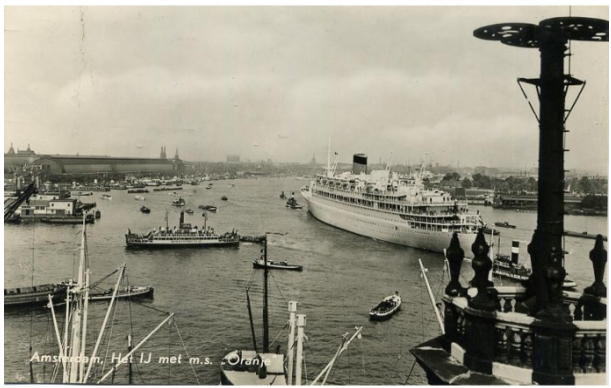

**Figure 3.** (*top*) KNMI building in Amsterdam, showing the time-flap system on top of the left-most tower (*bottom*) Close-up (postcard). (Klaus Hülse collection.)

Chronometers had to be wound regularly—ideally at the same time each day, shortly after noon, with the same number of turns, so that they continued to give correct readings (e.g., de Grijs and Jacob, 2021). It was a common property of mechanical chronometers that they would gradually gain or lose time each day. Master clockmakers understood that it was neither possible nor necessary to construct timepieces that were characterised by a rate of zero, that is, displaying the exact GMT at all times. Their overarching aim was to construct chronometers exhibiting rates that were as near constant as possible so that calculations of their longitudes at sea remained straightforward and only required a simple correction for the chronometer's drift.

By the late nineteenth century, Amsterdam's system had evolved beyond the classic falling ball. The prominent 'KNMI building',[3] constructed in 1884 at the Oostelijke Handelskade, featured meteorological instruments and a dedicated time-signal mast (see Figure 3). Initially, a time ball had formed part of the scientific equipment; it was located on the building's western tower (Willemsen et al., 2013: 43). Later, four large black time flaps (discs), each one metre in diameter and mounted some 31 metres above the ground, replaced the traditional ball. Rotated from horizontal to vertical five minutes before the signal and returned to rest at the appointed time, the flaps were more visible against the skyline and more mechanically reliable than the older balls. This installation remained active from at least 1898 until well into the 1930s (Kinns, 2022). Operational time-flap systems appear to have been confined to the Netherlands and her colonies, Belgium and northern France only; in Poland a similar approach using shutters was used (Hülse, 2001–2025; Kaiser, 1867: 21).

From 1928, Amsterdam added time lights, switched on five minutes before and off at the signal time. These lights made the signal visible even in poor weather or at night, reflecting the broader interwar transition to electrical systems. Flaps were retained as a backup in case of light failure until at least 1939, but by 1947 only the lights remained in use (Kinns, 2022). The KNMI building, which also housed the harbour authorities, customs, post and telegraph offices, was demolished in 1975. Its absence left a conspicuous gap on the waterfront until the *Muziekgebouw aan 't IJ*, Amsterdam's main concert hall for contemporary classical music, rose on the site at the beginning of the twenty-first century (Willemsen et al., 2013: 43).

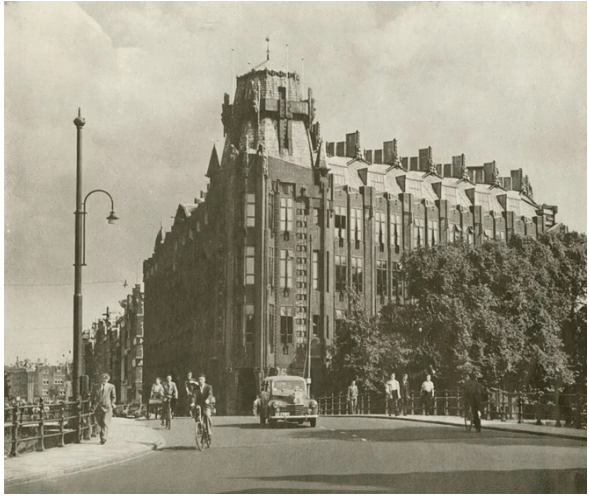

**Figure 4.** Amsterdam's Scheepvaarthuis, displaying a time ball on its tower. (Nieuwe Instituut, Amsterdam; public domain.)

Amsterdam's maritime quarter also hosted other time-signalling devices. The *Scheepvaarthuis* ('Shipping House'; Prins Hendrikkade 108), now the Grand Hotel Amrâth, was once equipped with a time ball mounted on its roof (see Figure 4). It was located at 52°2.487′ N, 4°54.229′ E (Brouwer, 1862). In the early twentieth century, when the building served as a headquarters for major shipping companies, the ball embodied the convergence of commerce, navigation and modernity. Although the ball is no longer present, the building survives as a Dutch national monument, with its maritime ornamentation recalling the city's shipping heyday.

Amsterdam thus presents a layered history of time signals. From an early naval-style ball in the 1840s, through the adoption of a more reliable time-flap mechanism in the late nineteenth century, to the modernisation of illuminated time lights in the interwar years, the city's harbour always maintained a visible public signal.

### 3.2 Nieuwediep/Willemsoord (Den Helder)

If Amsterdam represented the commercial face of Dutch maritime power, the Nieuwediep (the main canal, literally 'new deep') at the northern sea port of Den Helder embodied its naval counterpart. Since the Napoleonic era, the Den Helder base had served as the principal station of the Dutch Royal Navy, housing shipyards, guard ships and the naval dockyard at Willemsoord. It was therefore no surprise that the Nieuwediep became one of the first Dutch harbours to host a time signal.

Den Helder's time ball began modestly. As we saw in Section 2.2, in January 1846 the Minister of the Navy had authorised a provisional arrangement: on one of His Majesty's ships, initially the *Juno* but soon transferred to Z.M.S.[4] *Castor* (Pilaar and Obreen, 1846), anchored either in the Nieuwe Diep or at the Texel Roads,[5] a ball of roughly one Dutch ell (68–70 cm) in diameter would be hoisted and dropped precisely at noon local mean time. The routine followed standard Dutch practice. Officers were instructed that "the moment the ball *begins* to fall shall be taken as the instant of local mean noon".

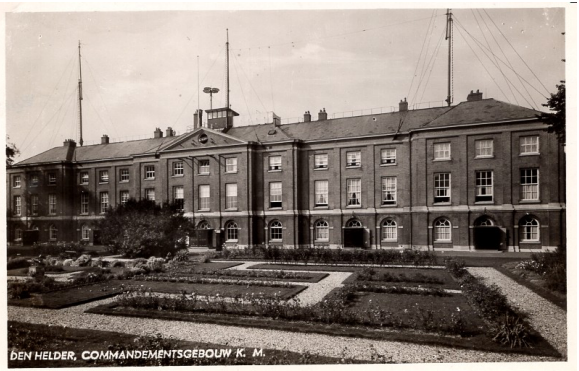

**Figure 5.** Time-flap system on top of the Dutch navy's administration building in Willemsoord, Den Helder. (Klaus Hülse collection.)

These provisional ship-based signals soon gave way to a permanent installation at Willemsoord, where an observatory, precision clocks and trained naval officers ensured regular operation. The ball was dropped at 11h 40m 53.8s GMT (Groustra, 1888: 161). It was located at 52°58′0″ N, 4°46′36″ E; the official longitude offset with respect to Greenwich was 19′ 7.1″ (Brouwer, 1862: 262; Brutel de la Rivière and Backer Dirks, 1859).

Detailed instructions survive: observations were made with an artificial horizon (Dekker, 1990; Hoefer, 1887: 62; Kaiser, 1860: 11), using a series of ten consecutive altitudes, discarding any that differed by more than one second. The command "stop!" was given one second before noon so that the ball's visible fall coincided exactly with the astronomical moment (Hondeijker, 1853: 171; Pilaar and Obreen, 1846). Naval manuals stressed that no chronometers should be carried to the observing site, lest they be disturbed; instead, a reliable pocket watch was compared before and after the observation.

By the mid-nineteenth century, Den Helder's system, like those of other Dutch ports, adopted the time flap in preference to the traditional ball (Kinns, 2022; see Figure 5). The new system was inaugurated on 1 January 1855 (Hoefer, 1887: 108; *Staatscourant*, 1855, Nos. 16 and 22). Its operation followed established practice. Four black discs, rotated from horizontal to vertical positions five minutes before the signal, providing a more visible and instantaneous cue. From a distance, the flaps resembled a disappearing time ball. The installation remained active well into the twentieth century.

The site's heritage value remains visible today. Much of Willemsoord has been preserved and redeveloped, now housing the Navy Museum and Marine Rescue Museum 'Dorus Rijkers'. Visitors can still see the monumental brick buildings and dry docks where the time ball once operated. Although the signal mast has not survived, historic photographs and technical drawings testify to its former presence, and the museum preserves related navigational instruments.

The combination of provisional ship-mounted balls, permanent dockyard apparatus, later time-flap mechanisms and surviving dockyard heritage illustrates the layered character of Den Helder's signals. Unlike Amsterdam, which reflected commercial needs, Den Helder's system was explicitly tied to the routines of the Dutch Royal Navy and the country's Hydrographic Service. Dutch survey vessels bound for its colonies relied on these signals to ensure their chronometers were trustworthy. In this sense, the Willemsoord time signal was not merely a local convenience but a piece of strategic infrastructure, connecting naval astronomy to imperial navigation.

Together, Amsterdam and Den Helder formed the twin poles of the emergence of Dutch time signalling—one commercial, the other naval.

### 3.3 Vlissingen (Walcheren)

Further south, the port of Vlissingen on former Walcheren island hosted a prominent time ball, with its drop timed at 12:00 local mean time (11h 45m 36.8s GMT; Ministerie van Marine, 1884: 34). Positioned at the mouth of the Scheldt River, gateway to Antwerp and a naval station, Vlissingen served both Dutch and Belgian vessels. From 1846 to 1855, the time-ball installation was positioned on the town's guard ship (Braat, 2010; Ministerie van Marine, 1884). It was then moved to the large warehouse of the Royal Naval Equipment Yard (at 51°26′23″ N, 3°35′17″ E; Brouwer, 1862: 262) until May 1877, before being installed on a stone tower adjoining the lock keeper's house at the western side of the main canal-to-sea lock (Ministerie van Marine, 1884), at 51°26.6′ N, 3°35.8′ E (*Schuttevaêr*, 1929).

By the middle of the nineteenth century, the Vlissingen time signal had been updated to consist of four black circular iron-mesh, disc-shaped flaps mounted on the arms of a horizontal cross atop a vertical standard (e.g., Hoefer, 1887: 108; Kaiser, 1860: 13–14), standing 19.5 metres above ground. It was clearly visible, at least to those armed with a pair of binoculars, from as far afield as the church tower in the provincial capital of Middelburgh (approximately 7 km in a direct line; 'Hbl.', 1854). Operating following standard Dutch practice, the mechanism soon encountered difficulties:

> My intention was that the signaller, by pressing a lever, could instantly move the signal flaps from the vertical to the horizontal position, or *vice versa*, and according to the model this seemed quite feasible. However, when implemented on a large scale, difficulties arose. The signal flaps were therefore moved by a heavy rod. Five minutes before local mean noon, they were brought into a vertical position, as a warning signal, by sliding the rod, and precisely at local mean noon the rod was released by a suitable mechanism, so that the flaps dropped under their own weight, instantly assuming a horizontal position. Care was taken to ensure that the rod could not fall with excessive force, and that the entire mast, with its signal flaps, could be gently lowered whenever the axles of the boards required cleaning. (Kaiser, 1860: 14; own translation)

The system's operation was calibrated by a pendulum clock housed in the same building and regulated twice weekly via telegraphic time signals from the Leiden Observatory (Hoefer, 1887; Kaiser, 1867: 21; Ministerie van Marine, 1884). The Leiden astronomers were particularly careful to adjust their measurements for differences in the ambient temperature, for which they derived a straightforward correction:

$$g = a + b(t - m) + c(t - m)^2,$$

where $g$ is the desired going rate, $t$ the instantaneous temperature, $a$ the chronometer's measured rate and $b$ its change per degree of temperature (°C), for a temperature $m$ (Kaiser, 1867: 22); $c$ is an adjustable parameter.

Archival records from 1898 indicate that Michiel Adriaan de Ruijter (1851–1923) held the position of *opzichter tijdbal* (time-ball supervisor; Zeeland Archives, 1898), thus confirming that dedicated personnel were employed to ensure the accuracy of the signal (even though it was no longer a ball by that time, despite De Ruijter's official title). This responsibility was critical: any timing error could compromise many ships' chronometers. The official recognition of such roles underscores the professionalisation of maritime time signalling in Dutch ports. Vlissingen's time signal was officially discontinued by 17 August 1929 (*Schuttevaêr*, 1929; see also Kinns, 2022).

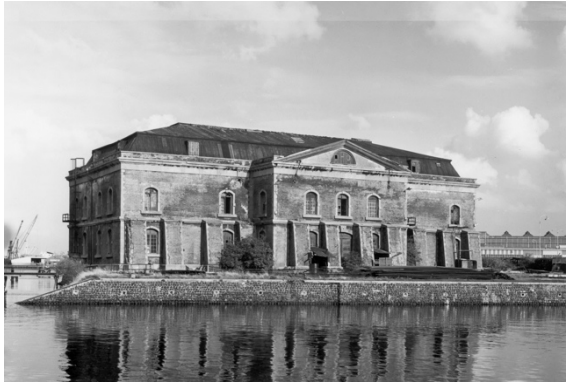

**Figure 6.** Arsenal building at the Vlissingen Royal dockyard. (Zeeuws Archief, Gemeente Archief Vlissingen, fa28386; public domain.)

The Vlissingen time signal was closely linked to meteorological observations. From about 1850, weather measurements—including barometric pressure, temperature, relative humidity, precipitation, visibility and wind direction—were recorded several times daily aboard the guard ship and later at the naval establishment. These data were submitted to the Dutch Meteorological Institute (KNMI) at the end of each month, forming the early basis for systematic meteorological observation in the region. By 1854, a time telegraph providing the mean local time had been installed on the roof of the Arsenal of the Royal dockyard (Hoefer, 1887: 108; see Figure 6, without telegraph equipment). It became operational on 1 June that year, to support both chronometer verification and accurate weather recording, with wind force measurements added in 1857 and storm warnings from about 1860 based on measurements with an *aëroklinoscoop*, a storm signal mast (Braat, 2010).

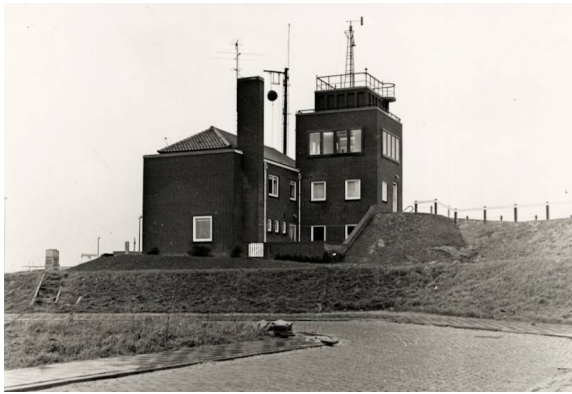

**Figure 7.** Vlissingen time ball on top of the lock keeper's building at the main canal-to-sea lock. (Klaus Hülse collection.)

The Vlissingen time ball, once mounted on a stone tower next to the main canal-to-sea lock (see Figure 7), was a distinctive feature of the harbour skyline. Its four circular plates would have been clearly visible from ships in the Scheldt estuary. In combination with nearby landmarks—including the Grote Kerk, the Roman Catholic Church, the Oranje- and Nieuwe Molen windmills, and the round tower by the Nieuwe Molen—the time ball served both as a navigational aid (Ministerie van Marine, 1884) and as a symbol of Vlissingen's maritime and scientific engagement in the nineteenth century.

### 3.4 Hellevoetsluis (Voorne)

Hellevoetsluis, located on the former island of Voorne, south of Rotterdam, also acquired a time ball. Once a home port of the Dutch East India fleet, Hellevoetsluis remained an important naval anchorage in the nineteenth century. The establishment of a time ball here illustrates the degree to which even secondary naval ports were integrated into the national system of maritime time signalling.

The port lies at the mouth of the Haringvliet inlet, providing access to Rotterdam and the southern Dutch rivers, strategically positioning it for both naval and commercial shipping. The time ball thus served a dual constituency, reflecting the hybrid military–commercial character of Dutch maritime activity. The Hellevoetsluis time ball was initially installed aboard a designated guard vessel, the Dutch Royal Navy frigate *de Schelde* (Hoefer, 1887: 107; Tindal and Swart, 1847: 160, 163). Each day at precisely noon local mean time (11h 43m 29.3s GMT), the ball, approximately one Amsterdam ell (69 cm) in diameter, would fall from the height of the main topgallant to the main mast topsail yard, with the operation following standard Dutch practice.

In 1855, an updated time signal, now equipped with flaps (discs), was installed on top of the town's hospital (Hoefer, 1887: 108; Kaiser, 1860: 14; see Figure 8), at 51°49′26″ N, 4°07′44″ E (Brouwer, 1862: 262). Observations for time determination were made using sextants with artificial horizons, thus ensuring precision to within a second (Hoefer, 1887: 107), and the time difference relative to Greenwich was noted as 16′ 32″. Its efficacy depended on careful shipboard preparation: chronometers had to be stable, shielded from temperature shifts and wound consistently. Although minor disturbances aboard a vessel could never be fully eliminated, these measures ensured that the fall of the time ball provided reliable verification of shipboard chronometers, a critical task for navigation in the busy waterways near Rotterdam.

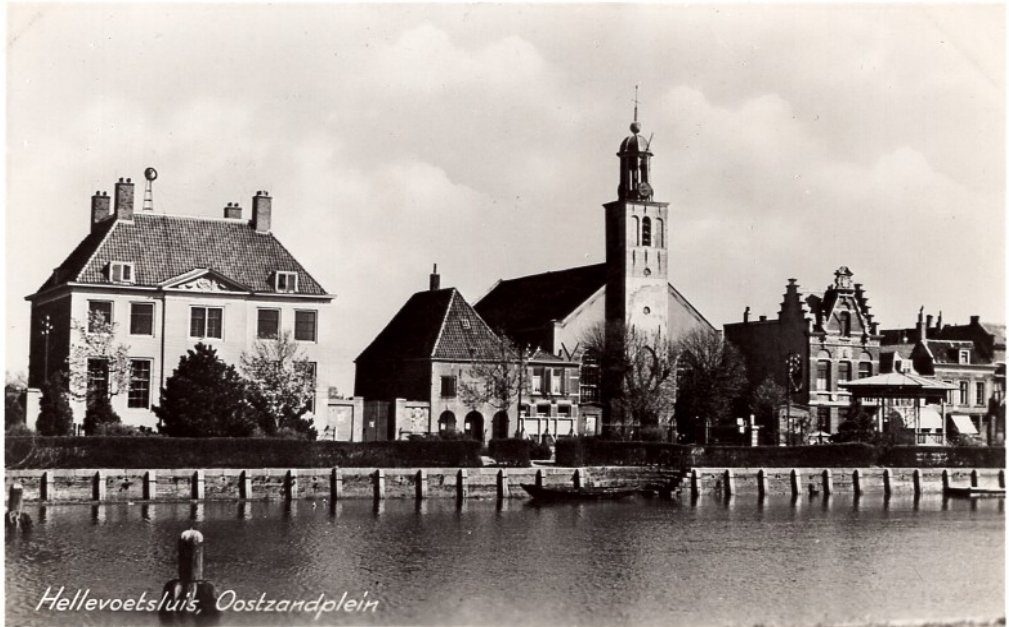

**Figure 8.** Time ball, Hellevoetsluis. (Klaus Hülse collection.)

### 3.5 Rotterdam

By the late nineteenth century, Rotterdam had overtaken Amsterdam as the country's main commercial port. Its time ball, with a fall timed at 12:00 mean local time (11h 42m 04.6s GMT), was therefore of great significance for merchant shipping. Rotterdam's rapid growth as a hub of transatlantic trade and as a centre for emigration to the Americas ensured a constant flow of vessels requiring reliable chronometer verification.

Rotterdam's system was tied to the city's meteorological and chronometric infrastructure. According to a contemporary report, the time was established at the Meteorological Institute at the *Poortgebouw* (Stieltjesstraat 38) using a chronometer made by the Amsterdam-based Danish clockmaker Andreas Hohwü (1803–1885). The system was synchronised twice weekly, on Tuesday and Friday evenings at 10 pm, with telegraphic time signals from the Leiden Observatory (Hoefer, 1887: vii, viii, 108; *Rotterdamsch Nieuwsblad*, 1897). GMT, reported from Leiden, was then converted to local Rotterdam time, with the city's longitude adopted as 4°30′ E for practical purposes. Differences in time between Amsterdam and Rotterdam, ranging from 1 min 33 s to 1 min 44 s, were formally accounted for in calculations for navigation and chronometer verification.

As in other Dutch ports, by the early twentieth century the original time ball in Rotterdam was replaced by a system of four black rotating flaps (see Figure 9), which were operated following standard practice.[6] In 1880, the system was listed as being located on the 'Tower of the Royal Dutch Yacht Club' at 51°54′30″ N, 4°28′51″ E, elevated 27 metres above ground on the south bank of the Nieuwe Maas (New Meuse).[7] By 1898 the flaps had been relocated to the Gate Building on the river's north bank (51°54′39″ N, 4°29′47″ E), about 1.1 km WNW of the previous site. The diameter of the discs remained 0.8 metres. By 1915, the installation was replaced by time flaps, moving once again, now to a position on the Rotterdam water tower on the north bank of the river at 51°54′23″ N, 4°27′01″ E, roughly 3.3 km ESE of the Gate Building (Kinns, 2022). From a distance, the rotating flaps simulated the appearance of a disappearing time ball, thus maintaining both practical and visual continuity.

Time lights were introduced in Rotterdam in 1921, illuminated five minutes before the signal time when the lights would be switched off. The flaps were retained as a backup until at least 1939; by 1947, only the lights remained (Kinns, 2022). This evolution illustrates the shift from mechanical to optical signalling, reflecting broader technological developments in maritime time dissemination.

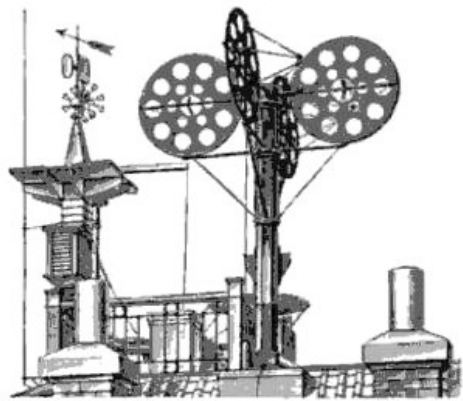
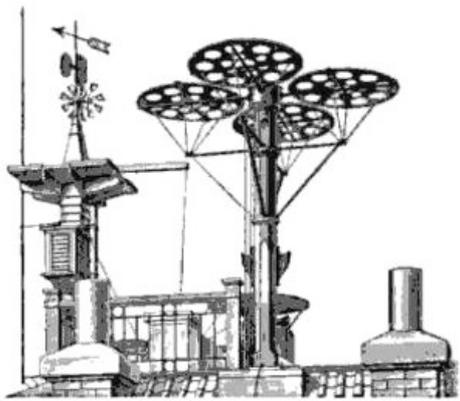

**Figure 9.** Time-flap operation in Rotterdam. (Zeitsignale für die Seeschiffahrt, *Meyers Konversations-Lexikon*, 6. Aufl. Beilage, 1909; out of copyright)

Rotterdam's time ball and its successor time flaps were a prominent feature of the riverfront skyline, particularly from the north bank of the Nieuwe Maas. Elevated on water towers or the Gate Building, they were designed for visibility across the harbour. The successive relocations of the flaps and later the introduction of time lights reflect both urban development and technological progress.

## 4 INSTITUTIONS AND STANDARDISATION

The network of Dutch time signals did not operate in a vacuum. Behind each daily time-ball or flap operation stood an intricate web of institutions, ranging from naval hydrographers and professional astronomers to municipal supervisors and telegraph operators. The authority of the signal depended not only on the visibility of the ball but on the confidence that it fell at the *correct* moment—confidence that could only be secured through institutional oversight, scientific legitimacy and integration with international standards.

### 4.1 International standards

From the early nineteenth century onwards, the Dutch navy assumed responsibility for maintaining international standards related to hydrographic surveying, nautical chart production and chronometer calibration (e.g., Mohrmann, 2003). Time balls naturally fell within this domain. They ensured reliable chronometers, indispensable to hydrographic work.

In this context, the Dutch Royal Navy was charged—in accordance with the new international standardisation—with charting the Dutch East Indies, implementing the Greenwich prime meridian on nautical charts, maintaining lighthouses, buoys and beacons, and publishing nautical information and notices. Merchant shipping could freely use these government provisions, including the time signals, and *Notices to Mariners* were published in the Dutch Government Gazette (Hondeijker, 1853: 173; Mohrmann, 2003).

In practice, the navy provided the officers who calculated local mean noon, ensured the correct raising and dropping of the ball (or operating of the flaps) and maintained the apparatus. The case of Vlissingen is illustrative. I already highlighted the role of the local time-ball supervisor, De Ruijter, in 1898. However, his role depended on the daily calculations and instructions issued by naval officers, who determined the precise offset between GMT and local mean time.

The Dutch Hydrographic Service also played a crucial role in publishing and disseminating the relevant temporal offsets from GMT. The navy's involvement ensured not only precision but also continuity. As the colonial files make clear (e.g., Kinns, 2021), Dutch naval officers were equally responsible for time-ball operation in Batavia and Paramaribo. The Dutch Royal Navy's global remit thus guaranteed that metropolitan and colonial practices remained consistent, embedding Dutch time balls within a broader imperial science of navigation.

### 4.2 Telegraphy as a disruptive development

If the navy provided the personnel and organisational framework, professional astronomers supplied the intellectual authority. The two most influential figures were Frederik Kaiser (1808–1872), professor of astronomy at Leiden University and director of its observatory, and Jacobus Cornelis Oudemans (1827–1906), director of Utrecht Observatory and later of the Indies Meteorological Service. Kaiser championed the use of telegraphic methods to transmit time signals from observatories to ports (see Section 5). Beginning in the 1850s, Leiden Observatory was connected to the telegraph network (Hoefer, 1887), allowing Kaiser to send precise time signals to railway companies and, potentially, to maritime installations. Oudemans went further, developing methods for determining longitude by telegraphic comparison of local times—a technique he applied in the Indies from 1871 onwards, linking Batavia to Singapore and Madras (present-day Chennai, India).

Their work illustrates the growing entanglement of 'pure' astronomy with practical navigation. Observatories that had once focussed primarily on celestial research now became nodes in the national infrastructure, generating signals that regulated railways, shipping and urban life. In the Netherlands, as elsewhere in Europe, the authority of the time ball derived ultimately from the authority of astronomy. When the ball fell in Amsterdam or Rotterdam, observers trusted it not because they could themselves observe the Sun crossing the meridian, but because they trusted that trained astronomers and naval officers had done so correctly.

The mid-nineteenth century witnessed a crucial technological innovation: the electric telegraph. For the first time, it became possible to transmit precise time signals over long distances instantaneously. In Britain, by the 1850s the Royal Greenwich Observatory began sending daily signals to ports and railway companies (Gillin, 2020). In the Netherlands, similar developments soon followed. By 1858, Oudemans had successfully used telegraphy to determine longitudes within the Netherlands (Mörzer Bruyns, 1985: 36). These methods were subsequently exported to the colonies, where telegraphic connections between Batavia, Singapore and Madras allowed the Indies Hydrographic Service to determine longitudes with unprecedented accuracy (e.g., Orchiston et al., 2021).

Telegraphy did not make time balls obsolete; it enhanced their authority. Whereas earlier balls relied on local astronomical observations, later installations could be synchronised directly with telegraphic signals from more distant observatories. The ball remained the visible public face of time, but its invisible backbone was now the telegraph.

Perhaps the most consequential institutional development was the adoption of Greenwich as the universal reference meridian (de Grijs, 2017: Ch. 7). For much of the eighteenth century, Dutch hydrographers had used Tenerife or Amsterdam as their prime meridian, in line with national traditions. By the mid-nineteenth century, however, all major Dutch ports had adopted Greenwich. This shift reflected both scientific and geopolitical considerations. Scientifically, Greenwich offered a stable and internationally recognised standard, increasingly indispensable for global navigation. Geopolitically, Britain's dominance in maritime trade meant that Dutch ships were effectively compelled to align with British practice. By timing their signals relative to Greenwich, Dutch authorities ensured compatibility with international navigation and reduced the risk of error.

The publication of precise offsets in nautical manuals underscored this alignment. Mariners in Dutch ports were instructed to compare their chronometers not to Amsterdam time *per se*, but to GMT as realised locally. In this sense, Dutch time signals became instruments of international standardisation, binding the Netherlands into the global regime of GMT. Manuals such as Jacob Swart's (*fl*. 1823–1866) *Verhandelingen en Berigten betrekkelijk het Zeewezen* (1847) circulated detailed instructions, embedding the time ball firmly within the professional literature of navigation.

In institutional terms, the Dutch system resembled those of other European maritime powers. In Britain, the Admiralty Hydrographic Office and the Royal Observatory at Greenwich oversaw time balls; in France, the *Bureau des Longitudes* and naval officers did likewise. The Dutch case, however, illustrates the adaptation of these models to a smaller maritime nation, although no less integrated. The relatively small scale of the Dutch system allowed for close coordination between the navy, professional astronomers and municipal authorities, thus ensuring accuracy without the need for the vast bureaucracies that characterised British practice (e.g., Baker et al., 2025). Without the global imperial network of Britain, the Netherlands focussed its resources on a few key metropolitan ports and two colonial outposts, Batavia (Kinns, 2021; Orchiston et al., 2021) and Paramaribo. The scale was smaller, but the institutional principles were identical: naval authority, astronomical legitimacy and international standardisation. The Dutch time ball or time-flap system was therefore more than a piece of harbour furniture. It was the visible tip of an iceberg of institutions, calculations and standards that linked the Netherlands to the wider world of nineteenth-century navigation.

**4.3 South of the border: Comparison with Antwerp**

Although Belgian, Antwerp was integral to the Low Countries' time-signal system. It was an early adopter of precision maritime time signals, reflecting the port city's commercial dynamism and industrial growth in the second half of the nineteenth century. The rapid expansion of European industry and shipping created a strong demand for accurate timekeeping in Antwerp, both for coordinating harbour labour and for regulating the departure of vessels along the Scheldt River.

By the mid-nineteenth century, this need could adequately be addressed through a combination of techniques, including time bars, time cannons and time flaps. Ultimately, a time-flap system with four round, horizontal discs was installed on the *Hanzehuis* (Hanseatic Merchant House) tower (see Figure 11). Raised five minutes before and dropped at 13:00, it followed British rather than Dutch practice.[8] The system was telegraphically connected to the Royal Observatory in Brussels, thus ensuring synchronisation with national time standards. In the event of a malfunction, a blue flag was flown on the mast for one hour to signal the failure (Ministerie van Marine, 1884; Willems, 2023).

The establishment of the Antwerp time signals was based on a longer history of local meridian observatories. By decree of King Leopold I (1790–1865), on 22 February 1836 the Belgian–French astronomer Adolphe Quetelet (1796–1874; Figure 12) had been tasked with standardising time across the newly independent Kingdom of Belgium. Among several meridian installations, a lesser-known meridian ran through the St. Pauluskerk (St. Paul's Church), intersecting with the small observatory on the Scheldt River where a meridian telescope was positioned. From 1837 to 1854, ships' captains arriving at Antwerp regularly calibrated their chronometers against this local time, thus ensuring that the official Antwerp time was consistently applied to both shipping and railway operations (Oyen, 2025: 34; Willems, 2023).

Visually, the time-flap mechanism at the *Hanzehuis* was prominent within the portscape (Ministerie van Marine, 1884) and clearly depicted in contemporary representations, such as Louis Van Engelen's (1856–1940) painting *Belgische landverhuizers* (1890; 'Belgian emigrants'; see Figure 13). The four discs, elevated above the docks, were designed to be visible across the harbour, providing a reliable reference for mariners while simultaneously signalling Antwerp's technological sophistication and maritime importance.

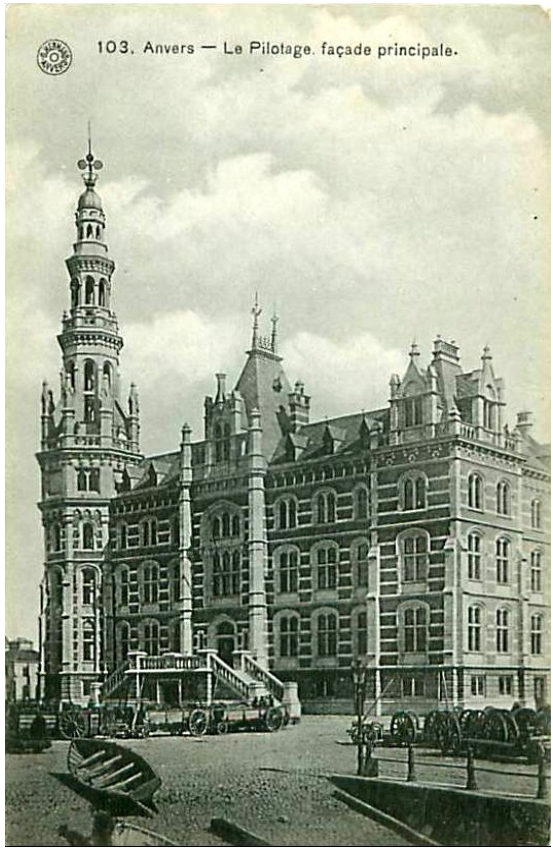

**Figure 11.** Time flaps on the *Hanzehuis*, Antwerp, ca. 1898–1939? (Klaus Hülse collection.)

The Antwerp time-flap system represents a distinctive variant of nineteenth-century European time signalling. Unlike Dutch ports such as Vlissingen or Rotterdam, which relied on falling balls or rotating flaps, Antwerp adopted the horizontal disc mechanism, integrating optical signalling with telegraphic precision. This combination of local observation, telegraphic connection to Brussels and visible mechanical cues exemplifies the ways in which commercial and industrial imperatives shaped the design and implementation of maritime timekeeping systems in the Low Countries.

The time-flap installation on the *Hanzehuis* tower remains a point of historical interest, illustrating the visible intersection of science, commerce and urban architecture. Its depiction in contemporary paintings, combined with archival descriptions, allows for a reconstruction of both its operational logic and its symbolic role as a public marker of temporal precision. Antwerp's early meridian observatories, particularly the St. Pauluskerk installation, further highlight the city's role as a pioneer in the standardisation of local time for maritime and railway traffic, a practice that preceded the widespread adoption of GMT.

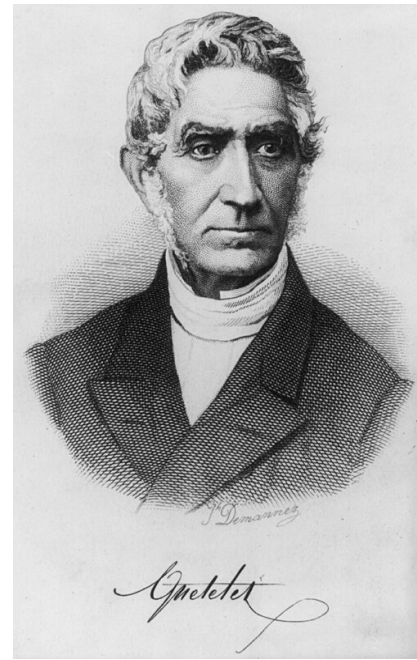

**Figure 12.** Adolphe Quételet (1796–1874) by Joseph-Arnold Demannez (1825–1902). (Steel engraving. Annuaire de l'Académie royale de Belgique, 41, 108 (1875); out of copyright.)

## 5 FREDERIK KAISER AND THE SCIENTIFIC AUTHORITY OF DUTCH TIME SIGNALS

The Dutch adoption of time balls cannot be understood without considering the interventions of Frederik Kaiser (Figure 14). Kaiser's reputation in the mid-nineteenth century was formidable: he was the leading public face of Dutch astronomy, a prolific writer and a figure consulted by ministers, princes and naval officers alike. His influence on time signalling extended far beyond Leiden. Through polemics in print, technical recommendations and practical innovations, Kaiser shaped both the apparatus and the legitimacy of the Dutch time-ball system.

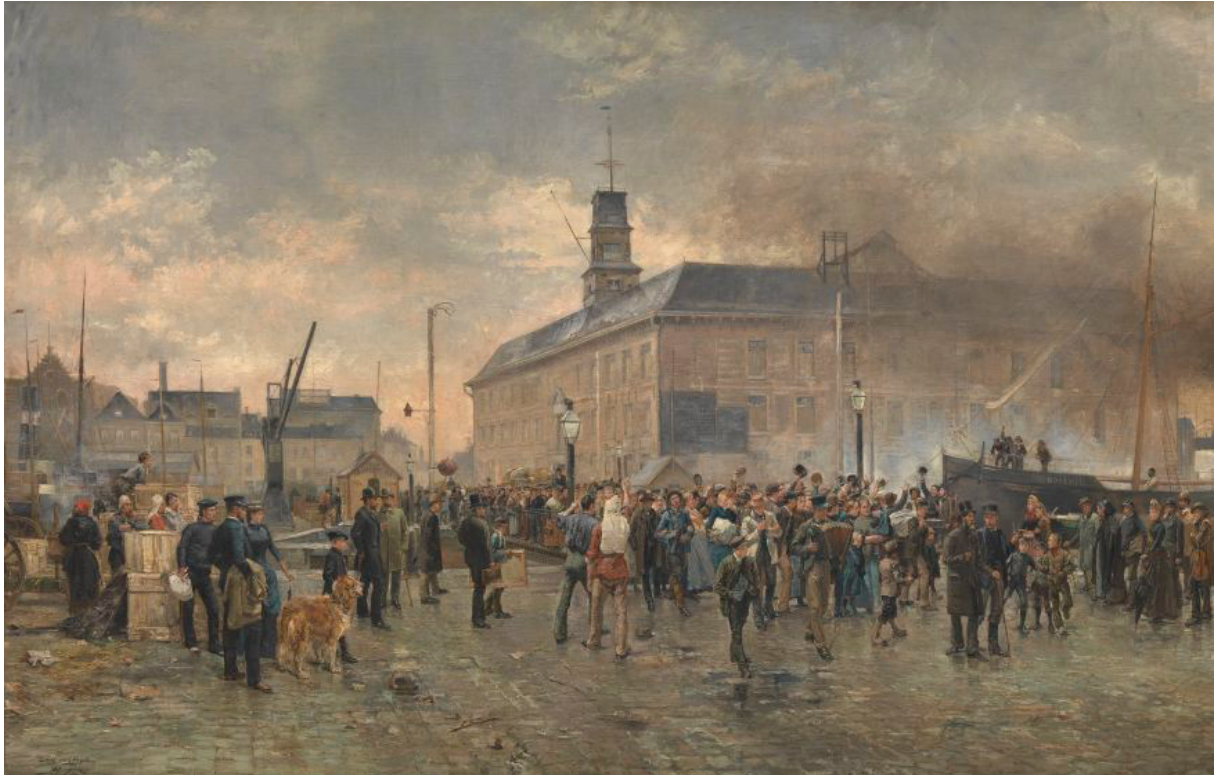

**Figure 13.** '*Belgische landverhuizers*' by Louis Van Engelen, 1890. (Vlaamse Kunst Collectie; CC0 1.0.)

### 5.1 Early debates and public polemics

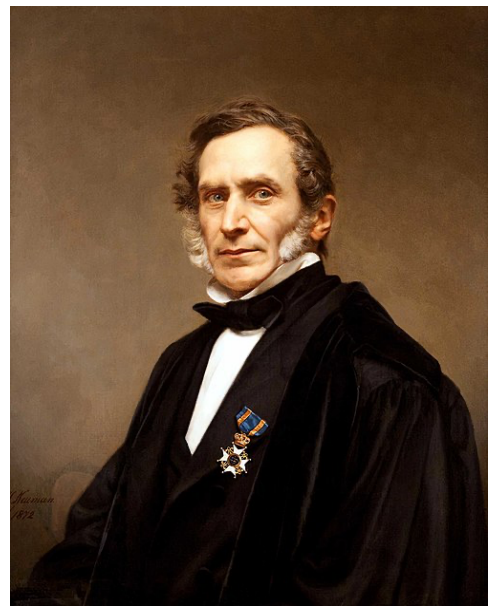

The first Dutch experiments with time signals at the Nieuwediep and Hellevoetsluis in the 1840s were controversial. Observers used sextants aboard guard ships to determine local noon and drop the ball. Critics charged that such determinations were too imprecise for chronometer calibration (e.g., Dekker, 1990; Hoefer, 1887: 107; Kaiser, 1860: 11). In 1846, an anonymous pamphlet argued that only a permanent land-based observatory with precision instruments could provide time accurate enough for navigation (Anonymous, 1846). Neither the Dutch university observatories nor Jacob Swart's astronomical chronometer observatory in Amsterdam could withstand the author's critical scrutiny (Dekker, 1990).

      Kaiser's response was characteristically forceful. His polemical tract, *Vermaning tot zwijgen* ('Warning to be silent'; Kaiser, 1846a), ridiculed the critic's claims and defended the navy's procedures. He insisted that sextant observations, if properly carried out, were sufficient, and that Dutch hydrographic officers were capable observers. Swart, custodian of nautical instruments, jumped into the fray and supported Kaiser in his defence, primarily relying on direct user experiences. For instance, one commander testified "…

**Figure 14.** Frederik Kaiser (1808–1872) by Johan Heinrich Neuman (1819–1898). (Painting, 1872; *Icones Leidenses* 240, Leiden University; public domain.)

that for about ten days he compared his chronometer with the drop of the ball at Nieuwediep and thereby obtained the same rate for his chronometer as he already knew from other observations" (cited by Dekker, 1990: 31). Swart thus concluded that the usual determination of time by sextant, as routinely carried out at the Nieuwediep and Hellevoetsluis, "… enjoyed the confidence of the competent evaluators" (Swart, 1847: 188). Nevertheless, Kaiser also seized the opportunity to highlight the broader deficiencies of Dutch scientific infrastructure. A well-funded system of small observatories, he argued, could provide chronometer-grade time at modest expense (e.g., Kaiser, 1846c).

This was not the only time Kaiser found himself embroiled in public controversy. He sparred with Swart over the virtues of balls versus flaps as signalling devices, and he debated the cost and design of the proposed observatories in Rotterdam and Amsterdam. His willingness to engage in print, often with sharp humour, ensured that the question of time signals was not confined to naval circles but became a matter of wider public discussion. From these debates emerged Kaiser's distinctive vision: a network of modest, inexpensive observatories linked to visible signals. Against those who demanded large, costly observatories (including the Dutch navy's leadership in both Vlissingen and Willemsoord!), Kaiser argued for economy and sufficiency. A small building with a pendulum clock, a transit instrument and a simple time-distribution mechanism would suffice (Hoefer, 1887: 108). Such stations, he suggested, could be placed in Rotterdam, Amsterdam and a series of key seaports, ensuring reliable signals without extravagant cost. Wrote Kaiser,

> Such an establishment must at the very least be equipped with a transit instrument and a pendulum clock. If a second pendulum clock is added, then in all investigations much greater accuracy can be achieved; and should one of the clocks unexpectedly stop, or need to be cleaned, another is immediately at hand, so that the work need not be interrupted. If one proceeds from the principle that the time observatory in question is to be established in a sea port, then in addition to the aforementioned instruments one should also provide certain important nautical instruments, such as: octant, sextant, the prismatic circle, various kinds of compasses—although their needles must be kept as far as possible from all timekeeping instruments—artificial horizons, and so forth. If one further adds a good telescope and some books and nautical charts, to be used in any nautical instruction given at the observatory, then the equipment may be considered adequate. (Kaiser, 1846b,c; own translation)

Kaiser also left his mark on the type of building one ought to make available—or undertake to construct if new-built—for his proposed observatory:

> If a building is already available for use, then a modest expenditure may suffice to adapt it into a time observatory; but if a new building must be erected, it could be constructed on a single storey, with a width of 17 feet [5.2 m] and a length of 38 feet [11.6 m]. The façade, facing either east or west, would have a central door and two windows. On one side, there could be a room 15 feet [4.6 m] long and wide, in which the chronometers could be kept. In one corner of this room, a pendulum or astronomical clock should be placed in a stone niche. A second room would serve the director of the establishment, while a small room behind these could house the transit instrument; for this purpose, the little room must be provided with a flat roof that can be opened, thereby offering the transit instrument an unobstructed view of the sky. (Kaiser, 1846b,c; own translation)

Continuing, he expanded,

> If local conditions allow, it is best to place the signal mast for the time signal entirely outside the building, because strong winds can shake the mast, potentially damaging the structure. The usual balls (time balls) that drop at a set moment should be avoided: they tend to hang or sway, and the start of their fall is difficult to observe precisely. It is far better to use flaps, which can be brought abruptly into a vertical position, as is done with railway telegraphs. Two pairs of flaps should be used, crossing each other, so the signal can be seen from all directions. Next to one of the pendulums, a lever can be installed, requiring only slight pressure from the signaller to bring the flaps instantly into the vertical position. The signal should be operated by a trained observer alone, without an assistant, and the mechanism must be designed to be completely reliable. (own translation)

## 5.2 From time balls to time flaps

This latter passage includes Kaiser's most striking proposal, that is, the replacement of time balls by signal flaps (e.g., Kaiser, 1860; Kaiser, 1867: 21). He argued that balls were flawed: friction delayed release and the descent was hard to observe. Flaps, by contrast, could snap into a vertical position instantaneously, leaving no ambiguity. They were also more visible against the sky, especially in poor weather (cf. *Algemeen Handelsblad*, 4, 5 and 10 July 1913), and could be constructed and maintained at lower cost. Wrote Kaiser,

> The customary balls, which drop at a given moment, must be rejected: firstly, because they too easily remain suspended, and secondly, because the beginning of the fall cannot be observed with sufficient precision. It is far better to make use of discs [flaps], which can be suddenly brought into the vertical position, as is the case with the telegraphs along the railways. Yet here one must employ two pairs of discs, crossing one another, so that the signal may be observed from all directions. (Kaiser, 1846c; own translation)

Not everyone was immediately convinced of the utility of time flaps, however. Wrote Swart in 1846, "With these ideas of Professor Kaiser concerning a chronometer observatory, we can fully agree. We would only not, for the time being, go so far as to replace the time balls with discs" (Swart, 1847: 153). Continuing, he explained, "… we would continue to prefer time balls as the time signal, especially on tall buildings: failure of the drop is hard to imagine, provided all reasonable precautions are taken, and with sufficient weight of the ball, the start of the fall can be clearly distinguished" (Tindal and Swart, 1847: 153–154; own translations).

However, by the early 1850s, Kaiser's advocacy bore practical results. New installations at Vlissingen, Willemsoord and Hellevoetsluis incorporated time-flap mechanisms, and models of his design were displayed in The Hague, the location of the Dutch Parliament. Although not all naval officers were convinced—Swart continued to defend the symbolic visibility of the ball—the fact that Kaiser's ideas were implemented demonstrates his influence over the apparatus itself.

Kaiser's role was not confined to the design of apparatus. He also mediated between metropolitan science and naval routine. Naval officers were responsible for daily operation, but Kaiser's imprimatur gave their signals credibility. When disputes arose over accuracy, the navy could point to Kaiser's involvement as evidence that their practices met scientific standards. The case of the Willemsoord observatory illustrates this dynamic. When, in early 1850, Prince Hendrik (Henry) of the Netherlands (1820–1879) donated a precision pendulum clock made by Hohwü to the navy in Willemsoord, questions arose as to how it should be incorporated into naval routines. Kaiser advised on its installation, ensuring that the clock's time was distributed to the harbour with the necessary precision (Kaiser, 1860: 12–13).

Kaiser thus became the guarantor of accuracy. Even when the mechanisms were not those he preferred, his participation reassured the public and the naval establishment that Dutch time signals were scientifically sound. Perhaps his greatest contribution was insisting on telegraphic time distribution. By the late 1850s, he had grown impatient with sextant observations at naval stations. Cloudy weather often prevented observations, and winter conditions in particular produced gaps. A more reliable method, he argued, was to distribute time directly from Leiden Observatory by electric telegraph:

> When I had communicated my thoughts on this subject to His Excellency the Minister [of the Navy, in 1858], he made me aware of a report, issued in June 1855 by the Honourable Jonkheer A. Klerck [1820–1876; Jonkheer: hereditary member of the untitled Dutch nobility], then Adjutant to the Director and Commander of the Navy at Vlissingen. As Adjutant, Jonkheer A. Klerck was responsible for two chronometers stationed at Vlissingen, as well as for giving the time signals. However, His Honourable Excellency found major objections in this arrangement, partly due to the lack at the time of an astronomical pendulum clock, and partly due to the impossibility, especially in winter, of obtaining regular time determinations with the sextant. Jonkheer A. Klerck proposed that, once a week, the time should be transmitted from Amsterdam to Vlissingen by the electric telegraph, with the cooperation of Mr. A. Hohwü. No action was taken on that proposal at the time, and indeed it was very doubtful whether a better determination of the time could be obtained in Amsterdam than at Vlissingen. (Kaiser, 1860: 16; own translation)

## 5.3 Leiden Observatory, centre of the Dutch time enterprise

This approach, with time signals broadcast from Leiden rather than Amsterdam, was implemented from 1859. Daily telegraphic signals were transmitted to the major Dutch seaports and the main commercial centres of Amsterdam and Rotterdam from 6 September that year (Hoefer, 1887: 108; Kaiser, 1860: 16–18; Kaiser, 1867: 21).

While Kaiser's imprimatur lent authority, the credibility of Dutch time balls also rested on the established routine of chronometer checks. Naval manuals and nautical periodicals spelt out procedures for observing the daily signal with the necessary rigour. Mariners were instructed not merely to note the instant of the ball's descent but to compare the interval between successive drops with the accumulated time shown on their chronometers. The difference revealed the instrument's going rate, while the divergence at the moment of the fall indicated its error (Hondeijker, 1853: 171).

To ensure reliability, observers were urged to repeat the exercise over several days and calculate a mean correction. Instructions even explained how to subdivide seconds by vocal counting: "… with a little practice, …" a sailor could count *one, two, three, four, five* between ticks of the chronometer's seconds hand, thus estimating quarter- or fifth-second fractions (Hondeijker, 1853: 171). Such details reveal a culture of precision in which Kaiser's insistence on accuracy was translated into embodied, almost ritual practice.

Published examples demonstrate the method in action. For instance, in the mid-1800s a chronometer aboard a vessel at the Nieuwediep was checked against five consecutive ball drops. The timepiece consistently gained between two and three seconds per day, yielding a mean rate of +2.55 seconds per day (Hondeijker, 1853: 171). At Hellevoetsluis, similar checks were recorded with another chronometer, whereas at Batavia the local time ball was treated in the same manner, its fall calculated against GMT by accounting for the longitude of the colonial port (Pilaar and Obreen, 1842). These examples not only illustrate the mathematics of correction but also the geographic reach of Dutch time-signal practice: the same procedures governed chronometer checks in the North Sea, the Maas (Meuse) estuary and the East Indies (Kaiser, 1860: 13).

The manuals also counselled mariners to note published particulars of each installation—the precise local mean time of the fall, the longitude relative to Greenwich and the procedures of hoisting and warning. Newspapers and maritime journals carried such information at the time of each ball's establishment, and seamen were advised to copy these notices into their logbooks. This documentary culture reinforced Kaiser's view that scientific time should be institutionalised and made reproducible. His advocacy for telegraphic distribution was, in effect, an extension of these prescribed practices to instantaneous transmission.

Kaiser developed a system by which signals determined at Leiden were transmitted daily to the naval stations at Willemsoord, Hellevoetsluis and Vlissingen, as well as to Amsterdam and Rotterdam.[9] Special receiving apparatus converted the telegraph clicks into audible beats synchronised with pendulum seconds. Kaiser's own 'secunden-klepper'—a device using the vernier principle to resolve fractional seconds—improved accuracy further (for details, see Hoefer, 1887: 109–110). This system profoundly reconfigured the authority of time signals. No longer were they based on local observations; instead, they were tied directly to the nation's premier astronomical institution. Leiden Observatory became the metronome of Dutch maritime life and Kaiser its conductor. The daily drop of the ball in Vlissingen or Rotterdam was thus an extension of Kaiser's calculations in Leiden, mediated by copper wires and telegraph keys.

**5.4 Kaiser's legacy**

Kaiser also pressed for the integration of telegraphic time distribution into the emerging railway system. In reports to the Minister of the Interior, he argued that trains should be regulated by signals from Leiden (Kaiser, 1855: 75; Kaiser, 1860), thus ensuring consistency across the national railway network. His proposals anticipated the eventual unification of Dutch civil time around a single telegraphic standard. Although resistance and budgetary constraints limited immediate implementation, Kaiser's advocacy linked maritime signals to a broader discourse on temporal discipline. He insisted that accurate time was not only a naval necessity but a national resource, essential for commerce, transport and scientific credibility.

Despite his influence, Kaiser did not win every battle. Many naval officers valued the symbolism and spectacle of the falling ball, which was visible from afar and attracted public attention. Flaps, although more precise, lacked the same dramatic effect. For Swart and others, the ball was both a navigational tool and a civic symbol, binding town and harbour in a shared moment of noon. Kaiser acknowledged this but remained sceptical. For him, science demanded precision above all. He repeatedly stressed that the signal must be unambiguous and that the needs of mariners outweighed considerations of spectacle. His writings reveal a persistent tension between scientific precision and public symbolism.

By the time of his death in 1872, Kaiser had left an indelible mark on Dutch time services. He had defended the navy against its critics, proposed alternative apparatus, established telegraphic distribution and linked maritime signals to wider debates on railways and civil time. His interventions ensured that Dutch ports kept pace with international standards, even as Britain and France expanded their networks.

Kaiser's paradoxical stance—sceptical of time balls yet crucial to their legitimacy—highlights the complex interplay between science and technology. The Dutch time-ball system owed its credibility to a figure who doubted the apparatus itself but insisted on the authority of the signals. Without Kaiser, the balls might have remained provisional naval contrivances; with him, they became emblems of Dutch scientific modernity.

Kaiser's story also clarifies the relationship between institutions and individual authority. Section 4 highlighted the role of the Dutch Royal Navy, the Hydrographic Bureau and professional astronomy in establishing Dutch time signals. Kaiser embodied these connections in a single person: a university astronomer whose influence extended into naval practice, government policy and civic life. Whether advocating for cheap observatories, defending naval officers or connecting Leiden to the ports by wire, Kaiser exemplified the role of science in the making of modern temporal infrastructure.

## 6 RECEPTION AND USE

The significance of Dutch time signals cannot be understood without considering their reception among those who used and observed them. Over time, they acquired symbolic weight as emblems of scientific modernity, national authority, and imperial reach. Their eventual decline in the age of telegraphy and wireless further shaped how they were remembered and commemorated. In this section I examine these dimensions in turn.

### 6.1 Practical reception by mariners

For navigators, the daily signals were indispensable. Since the late eighteenth century, marine chronometers had transformed longitude determination at sea. Yet even the best instruments were imperfect. They drifted, sometimes unpredictably, and needed to be corrected regularly by establishing their going rate. The time balls in Amsterdam, Den Helder, Vlissingen, Hellevoetsluis and Rotterdam provided just such standards.

The procedure of observing the drop was both simple and exacting. A ship's officer, often the navigator or first mate, stationed himself with a clear line of sight to the ball. He compared the exact moment of the drop against the chronometer's indication, recorded the discrepancy and then used this correction to calculate longitude. In principle the operation lasted seconds; in practice it demanded discipline, attention and favourable conditions. If fog obscured the ball, or if the vessel was not in harbour at the appointed hour, an opportunity was lost until at least the following day.

The consequences of inaccuracy were serious. A single second's error in time corresponded to a quarter of a nautical mile in longitude at the equator. If the chronometer drifted by ten seconds over a long voyage, the ship's position could be off by more than two nautical miles—enough to spell danger when approaching a coast or reef (e.g., *Algemeen Handelsblad*, 4, 5 and 10 July 1913). Mariners therefore regarded the Dutch signals not as conveniences but as safeguards, as vital to safe passage as charts or soundings. Hydrographic publications repeatedly stressed this point, urging captains to take advantage of time signals whenever possible.

Dutch seafarers were not unique in this reliance. British, French and German mariners followed similar practices. Yet the Dutch case is notable for the density of its signalling network relative to the country's size and the integration of naval and commercial ports. A vessel departing from Amsterdam could check its chronometers at the Westerdok departure point, again at Hellevoetsluis or Vlissingen if sailing southwards, or at the Nieuwediep on a northbound course, each time refining its going rate. The redundancy was deliberate: it maximised opportunities to catch errors before they could compound.

Mariners were also acutely aware of the limitations of the system. Mechanical failure of the ball's release mechanism occasionally delayed the drop, and observers sometimes disagreed about the precise moment when the ball began to fall. These challenges reinforced the desirability of multiple observations. Navigators often recorded not just a single sighting but several successive days of drops, from which they could average out anomalies.

For Dutch naval officers, the signals also held institutional significance. They formed part of a larger system of discipline and routine, binding the fleet to a common temporal rhythm. Ships in port calibrated their instruments simultaneously, creating uniformity across the squadron. This synchronisation was not only technical but social, too: it expressed the Dutch navy's collective adherence to precision and order.

Although they were not prepared to (co-)fund the time signals (Dekker, 1990),[10] insurance companies, too, recognised the importance of chronometer calibration. The presence of a time ball in a port reduced the risk associated with a voyage, thereby lowering premiums. In this sense, time signals contributed indirectly to the competitiveness of Dutch shipping. Rotterdam's rapid rise as a commercial hub in the late nineteenth century owed much to such infrastructural improvements, which reassured ship owners and insurers alike that Dutch ports offered modern facilities comparable to those of London or Hamburg.

## 6.2 Civic spectacle and urban life

If mariners approached the time ball as a scientific instrument, townspeople experienced it as a civic performance. The gradual ritual provided a daily drama. Crowds often gathered near the site, especially in Amsterdam and Rotterdam, to watch the standardised sequence unfold. For those without watches, the ball was one of the few reliable public signals of exact time. Even for those who did own timepieces, it provided a chance to check and reset them.

The symbolism of regularity mattered as much as the information conveyed. The drop punctuated the urban day with mechanical precision, offering reassurance that the rhythms of commerce, industry and transport were anchored to a trusted standard. It complemented other temporal markers such as church bells, factory whistles and railway timetables, but unlike those it made the scientific character of time visible.

Postcards provide vivid evidence of how the time balls entered popular visual culture. In the late nineteenth and early twentieth centuries, views of Amsterdam's Westerdok and Rotterdam's harbours often included the mast with its prominent sphere. These images were not incidental; the time ball was a recognisable landmark, signalling to visitors that the city participated in the modern world of precise navigation and global commerce. Sending or collecting such postcards was itself a way of engaging with this identity, turning the signal into a token of civic pride.

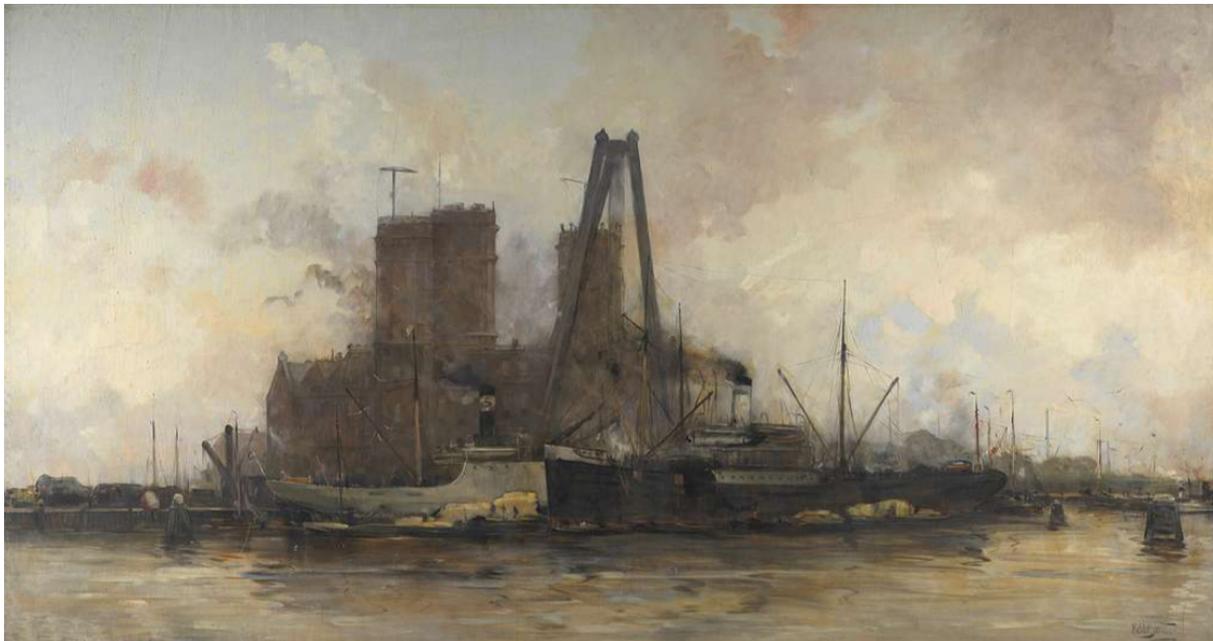

**Figure 15.** '*Gezicht op de kop van de Oostelijke Handelskade*' by Hobbe Smith (1862–1942). (Painting, 1913; Amsterdam Museum; CC BY-NC-SA 3.0).

Even more striking is Hobbe Smith's (1862–1942) 1913 painting of the Amsterdam time ball above the Westerdok, *Gezicht op de kop van de Oostelijke Handelskade* ('View of the tip of the Oostelijke Handelskade'; see Figure 15). It depicts the KNMI building in Amsterdam with its towers and signalling mast prominently overlooking the IJ waterfront. Smith, one of the city's most celebrated painters, chose the ball as a central motif in a scene otherwise dominated by shipping and harbour activity. When the painting was featured at the ENTOS exhibition[11] in Amsterdam-North that year, it was warmly received, and critics singled out its evocation of Amsterdam's maritime identity. By placing the ball at the centre of a work of high art, Smith underscored its cultural resonance. It was not only a tool for sailors but an emblem for citizens, a sign of belonging to a seafaring nation attuned to the latest scientific practices (Willemsen et al., 2013: 43).

Unlike the invisible signals of telegraphy that would later replace it, the time ball was meant to be seen. Its descent drew collective attention, if only for an instant, to a shared standard of measurement. In an era when mechanical clocks were becoming common but not yet universal, such moments carried considerable social weight. They reassured townspeople that the city was ordered, disciplined and synchronised with the broader world.

### 6.3 Symbols of science, empire and modernity

Beyond their immediate function and spectacle, Dutch time signals acquired symbolic force as markers of science, empire and modernity. To raise and drop the ball each day was to enact the authority of astronomical calculation, institutional discipline and state power. The signals demonstrated that the Netherlands, though smaller than Britain or France, participated fully in the international regime of scientific navigation.

In this respect, the Dutch case paralleled but also diverged from others. At Greenwich, the time ball atop the observatory symbolised the British claim to global temporal authority, reinforced by the eventual adoption of GMT as the international standard. Paris had its own ball at the Palais du Luxembourg, while Hamburg, Copenhagen and Stockholm built similar installations (Kinns, 2022). Dutch signals were smaller in scale but no less significant for projecting an image of scientific credibility.

They also carried imperial implications. Dutch ships bound for the East and West Indies relied on chronometer calibration before their long voyages. By guaranteeing reliable time, the signals facilitated imperial expansion and control. The ability to sail accurately across oceans, to reach colonies efficiently and to return with cargoes safely was inseparable from the ability to measure time precisely.

The signals further materialised the notion that time could be measured, standardised and publicly displayed with precision. In doing so, they contributed to what historians of technology have called the 'disciplining of time' (Thompson, 1967; see also, e.g., Glennie and Thrift, 1996; Landes, 1983)—the process by which industrial and commercial societies internalised temporal regularity.

### 6.4 Failures, decline and memory

Like all technologies, time balls were eventually overtaken. The spread of the electric telegraph in the mid-nineteenth century and wireless signals in the early twentieth century made visible drops increasingly redundant. Telegraphy could transmit time instantaneously across great distances; radio could reach ships at sea directly. By the 1920s, the utility of time balls had largely faded, although some persisted into the interwar period for their symbolic value.

Yet the memory of the signals lingered. In Amsterdam, the mast above the Westerdok became an element of the cityscape preserved in art and postcards. At Den Helder, the former dockyard complex now houses the Dutch Navy Museum, which commemorates the naval role of the time signal. In Rotterdam, postcards continue to testify to the ball's former presence, even as the harbour modernised beyond recognition.

The endurance of these memories reflects the unusual duality of the time ball. As a scientific instrument, it was obsolete once telegraphy and radio offered superior precision. As a cultural symbol, however, it retained its power. The fall of the ball was etched into the routines of sailors and citizens alike, remembered as both practical aid and civic spectacle.

The reception and use of Dutch time signals thus operated on multiple levels. For mariners, they were indispensable tools for ensuring safe navigation through the calibration of chronometers. For citizens, they were reassuring spectacles that disciplined the rhythms of urban life and offered visual confirmation of scientific modernity. And for the Netherlands as a whole, they became symbols of participation in a global order of precision, empire, and rationality. Their eventual decline did not erase these meanings but transformed them into memory and heritage.

## 7 DECLINE AND TRANSFORMATION

The nineteenth century was the age of the time ball; the twentieth was the era of its obsolescence. Even as Dutch ports embraced the daily ritual of the ball's rise and fall, new technologies emerged that promised greater accuracy, wider reach and less dependence on the weather or visibility. The gradual decline of Dutch time balls was neither abrupt nor uniform. It unfolded over decades, shaped by the interplay of telegraphic innovation, wireless communication and shifting cultural expectations of time.

The first challenge to the time ball came from the electric telegraph. In Britain, the Royal Greenwich Observatory began transmitting telegraphic time signals in the 1850s, initially to railway stations and later to maritime ports (Gillin, 2020). These signals allowed for instantaneous communication of precise time over long distances, circumventing the need for visual observation. The Netherlands followed suit within a few years. These developments had profound implications. Where the time ball required clear skies, an unimpeded line of sight and physical presence in the harbour, the telegraph could deliver time to multiple sites simultaneously and without delay.

In practice, telegraphic signals did not immediately displace time balls. Rather, the two systems coexisted. The telegraph provided the authoritative reference, while the time ball served as its public manifestation. Mariners continued to observe the ball's fall, but increasingly they knew that the moment was set not by a local astronomer's meridian observation but by a telegraphic pulse transmitted from a more distant observatory. The ball remained visible, but its authority was now invisible.

The second—and ultimately decisive—challenge came with wireless telegraphy. From the early twentieth century, radio signals made it possible to broadcast precise time across vast distances, including to ships at sea. By around 1910, the Eiffel Tower was being used regularly to transmit daily wireless time signals (for example, midnight time signals), under the direction of the Paris Observatory and the *Bureau des Longitudes* (Baillaud, 1922; see also Howse, 1980: 155; Observatoire de Paris, 2015). By the early twentieth century, comparable wireless time-signal services were being developed in other countries: the Royal Observatory, Greenwich, began regular wireless transmissions in 1904 (Howse, 1980: 154–155), the German Norddeich radio station transmitted daily time signals from 1910 (Bureau des Longitudes, 1915) and the U.S. Naval Observatory initiated wireless time-signal broadcasts in 1904 (Bartky, 2000).

Globally, time signals were transmitted according to the same system: a continuous tone that would cease, after which, during the silence, the following words were spoken: "When the tone returns, the standard time is …" At that moment, the ship's chronometer could be read and the difference with the broadcast standard time established, which would then be entered into the chronometer log (Marinemuseum, 2015).

For the Netherlands, radio represented both an opportunity and a challenge. Dutch ships could now receive time signals directly at sea, reducing their dependence on port facilities. The Dutch Royal Navy quickly adopted wireless calibration methods, whereas merchant shipping companies also invested in radio equipment for their fleets. The impact on time balls was profound. Their *raison d'être*—providing a visible signal to calibrate chronometers in port—was gradually eroded as ships gained the capacity to calibrate at sea. By the 1920s, the time ball was no longer indispensable; by the 1930s, it was increasingly redundant.

Unlike in Britain, where the Greenwich time ball remains in operation as a heritage attraction, Dutch time balls gradually disappeared without ceremony. By the interwar period, most had ceased daily operation, and by the mid-twentieth century none remained functional. The process was uneven. Larger ports such as Rotterdam and Amsterdam discontinued their time balls relatively early, as shipping companies invested in wireless equipment. Smaller ports, where the cost of new technologies was higher and maritime traditions more conservative, retained their balls somewhat longer. Yet even here, the pressure of modernisation proved irresistible. Archival traces of this decline are fragmentary. Municipal records occasionally note the dismantling of a mast (e.g., *Schuttevaêr*, 1929: 18), or the reassignment of a time-ball supervisor. Hydrographic publications ceased to list fall times, signalling their obsolescence. By the 1950s, time balls had vanished from Dutch harbours altogether.

The Dutch experience was far from unique. Across Europe and the wider world, time balls disappeared in the early twentieth century, displaced by radio signals. Britain retained Greenwich as a heritage site, and a handful of colonial time balls—such as those in Sydney (Australia) and Port Elizabeth (South Africa)—survive as tourist attractions. In most cases, however, time balls were dismantled or left to decay. The decline of the time ball illustrates the broader trajectory of technological change. What was once cutting-edge infrastructure became, within decades, obsolete. The speed of this obsolescence testifies to the pace of nineteenth- and twentieth-century innovation. The time ball's rise and fall encapsulate the transition from visual, localised and communal time signals to invisible, global and instantaneous communications.

Although Dutch time balls disappeared physically, their legacy persisted in subtler ways. Dutch participation in international timekeeping organisations, contributions to geodetic surveys and integration into radio time-signal networks all built upon the foundations laid by the era of the time ball. In this sense, the time ball was not an endpoint but a bridge: a transitional technology that prepared Dutch society for the more abstract regimes of telegraphy, radio and eventually atomic time.

## 8 FINAL THOUGHTS

The Dutch adoption of time balls in the nineteenth century illustrates how a seemingly technical instrument could operate simultaneously as a scientific tool, a civic performance and a symbol of modernity. At one level, these signals served a precise and indispensable purpose: ensuring the accuracy of marine chronometers, the cornerstone of safe navigation and imperial expansion. Mariners depended on them to maintain their going rate, and the navy incorporated them into its wider routines of discipline and coordination.

At the same time, the daily rise and fall of the ball shaped the rhythms of urban life. Townspeople experienced the signals as spectacles of precision, visible enactments of science in the heart of the city.

Postcards, paintings and press commentary testify to their civic significance, projecting an image of the Netherlands as a nation both seafaring and scientifically progressive.

The dual character of the Dutch time balls—instrumental for mariners, emblematic for citizens—ensured that they left a lasting imprint. Even after telegraphy and wireless rendered them technically obsolete, they persisted in memory, preserved in art, heritage sites and cultural imagination. Their history thus reveals how science and technology can function not only as practical solutions to navigational problems but also as powerful public symbols. The Dutch case, modest in scale but rich in meaning, reminds us that the measurement of time was never only about accuracy: it was also about identity, authority and belonging in a modern world.

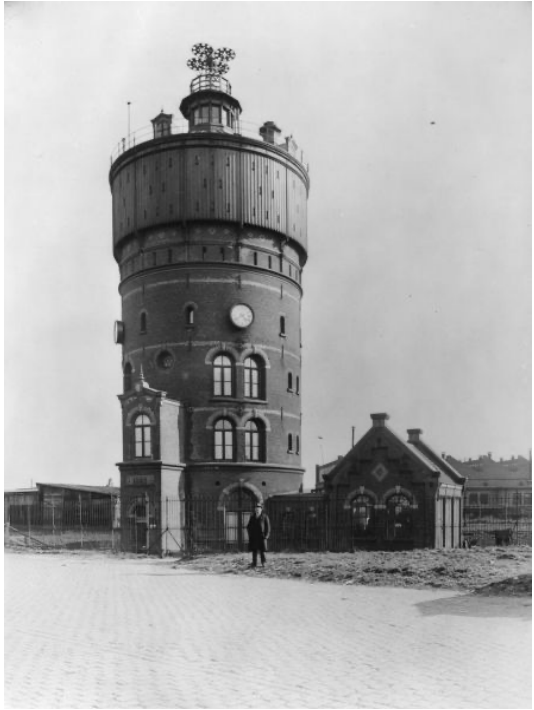

**Figure 10.** Water tower at the Delfshaven in Rotterdam showing its time-flap system, ca. 1922–1926. (Gemeente Archief Rotterdam; public domain.)

## 9 NOTES

1. The word *sjouw* is not usually associated with the meaning of time ball, as on Terschelling. In maritime usage, *sjouw* means a partially rolled-up or tied-up flag that is hoisted as a distress signal.
2. Some of the earliest maps featuring *laweis* include a map from 1664 of Schoterland (a former local government area in central Friesland) and a 1765 map used for peat extraction near Haulerwijk (eastern Friesland; van der Molen, 1978: 265). They were used until at least the turn of the twentieth century given that a *lawei* was still sold in 1896, in Hoornsterzwaag (southern Friesland).
3. KNMI is the acronym for the Royal Dutch Meteorological Institute, *Koninklijk Nederlandsch Meteorologisch Instituut* (contemporary spelling).
4. *Zijner Majesteits Ship* = His Majesty's Ship.
5. The main naval anchorage at Den Helder had long been the Texel Roads, a sheltered location off the coast of the northern island of Texel, just north of Den Helder.
6. A time *ball* appears to have been operational on the Rotterdam water tower in 1911, the only year it is listed (Kinns, 2022).
7. It is unclear whether this is the same location as Ruige Plaat, Delfshaven, where time flaps had been installed on a clearly visible tower (Hülse, 2001–2025): see Figure 10.
8. In 1833, John Pond (1767–1836), Britain's Astronomer Royal, designated 13:00 as the British reference time. This allowed the Greenwich astronomers to accurately verify their time calibration by carefully observing the Sun's meridian transit at midday and then drop the ball an hour later.
9. In 1887, the *Maatschappij voor Tijdaanwijzing* ('Society for Time Indication') appears to have had plans to establish time flaps and time balls of somewhat inferior accuracy (of order one second rather than one-tenth of a second) for the benefit of small vessels as well (Hoefer, 1887), but this idea does not seem to have materialised.
10. When the development of a full coastal time-signal system was first proposed, around 1850, both the Minister of the Navy—Vice-Admiral Engelbertus Lucas Jr. (1785–1870)—and the insurance sector balked at the costs involved, of order *fl*. 2000 (Algemeen Rijksarchief, 1850; Dekker, 1990; Dutch guilders), easily equivalent to a six-figure sum in Euros today.
11. The *Eerste Nederlandse Tentoonstelling op Scheepvaartgebied* was the First Dutch Maritime Exhibition.